\documentclass[letterpaper,english,aps,longbibliography]{revtex4-1}
\usepackage[T1]{fontenc}
\usepackage[utf8]{inputenc}
\usepackage{amsmath}
\usepackage{amssymb}
\usepackage{babel}
\usepackage{graphicx}
\usepackage{dcolumn}
\usepackage{bm}

\providecommand{\ket}[1]{\left\lvert #1 \right\rangle}

\begin{document}

\title{Non-Abelian Extensions of the Dirac Oscillator: A Theoretical Approach}
\author{Abdelmalek Boumali}
\email{boumali.abdelmalek@gmail.com}
\affiliation{Laboratory of Theoretical and Applied Physics~\\
Echahid Cheikh Larbi Tebessi University, Algeria}
\author{Sarra Garah}
\email{sarra.garah@univ-tebessa.dz ;
gareh64@gmail.com}
\affiliation{Laboratory of Theoretical and Applied Physics~\\
Echahid Cheikh Larbi Tebessi University, Algeria}

\begin{abstract}
We formulate the Dirac oscillator covariantly in the presence of external non-Abelian gauge fields.
More precisely, the matter field is written as $\Psi_{\alpha A}(x)$, where $\alpha$ denotes the Dirac index and $A$ the isospin index, so that the Hamiltonian acts on the tensor-product space $\mathbb{C}^{4}\otimes\mathbb{C}^{2}$ in the fundamental representation.
Starting from the gauge-covariant Dirac equation, we then implement the oscillator interaction through the standard non-minimal substitution and promote the construction to an $\mathrm{SU}(2)$ background.
In this way, we derive the associated non-Abelian field-strength tensor and isolate the commutator contribution, which has no Abelian analogue.
Consequently, the generalized Pauli interaction $\sigma^{\mu\nu}\mathcal{F}_{\mu\nu}$ produces matrix-valued spin--isospin couplings.
At the same time, the Abelian sector reduces to the conventional Moshinsky--Szczepaniak Dirac oscillator, whose exactly solvable spectrum provides a natural benchmark for the extended theory.
For an aligned planar background, the commutator term yields the closed-form isospin splitting
\begin{equation*}
E_{\pm,n,\ell,m_t}=\pm E^{(0)}_{n,\ell}-\zeta m_t,
\end{equation*}
thereby making the internal-Zeeman mechanism explicit.
To render this result visually transparent, we also include a direct spectral representation of the exact branches as functions of the splitting parameter $\zeta$.
Furthermore, we show that the planar Dirac oscillator admits a direct correspondence with effective graphene Hamiltonians: the same non-minimal confinement rule generates the graphene Dirac-oscillator form once the relativistic mass scale is traded for the appropriate gap parameter.
In turn, retaining the internal valley or layer doublet extends this correspondence to the gauge-field case, where matrix-valued effective connections provide a natural implementation of the non-Abelian model and the commutator term controls a possible lifting of internal degeneracies.
Overall, the resulting framework separates universal commutator-driven effects from background-dependent kinematic shifts and thus provides a controlled setting for studying relativistic bound states in Yang--Mills backgrounds and in graphene-based Dirac materials with effective non-Abelian structures.
\end{abstract}

\pacs{04.62.+v; 04.40.-b; 04.20.Gz; 04.20.Jb; 04.20.-q; 03.65.Pm; 03.50.-z; 03.65.Ge; 03.65.-w; 05.70.Ce}
\keywords{Dirac oscillator, non-Abelian gauge fields, relativistic quantum mechanics, Dirac equation, spin-orbit interaction, Lie groups, quantum field theory, SU(2) gauge theory, non-minimal coupling, particle physics, QCD, electromagnetic interactions}
\maketitle

\section{INTRODUCTION}
In the Abelian setting, the correspondence between the $(2+1)$-dimensional Dirac oscillator and graphene/graphene-like systems has already been explored through effective magnetic-field mappings, chiral graphene carriers, and confined graphene structures \cite{QuimbayStrange2013GrapheneDO,Mandal2010DiracOscillator,Belouad2016GateTunable,Boumali2015ThermodynamicGraphene}.
More broadly, analytically tractable relativistic models with nontrivial spin structure remain valuable as both conceptual laboratories and effective descriptions.
Among them, the Dirac oscillator occupies a distinguished place because it combines exact solvability with a rich pattern of finite and infinite degeneracies \citep{Moshinsky1989DiracOscillator}.
Accordingly, it provides a natural framework for examining how relativistic confinement, spin--orbit structure, and external fields interact within a mathematically controlled setting \citep{Moshinsky1989DiracOscillator}.
From this viewpoint, gauge interactions offer a systematic way to test the stability of the solvable Dirac-oscillator structure.
In Abelian backgrounds, the Dirac oscillator has also been explored in the presence of magnetic confinement and related planar Dirac-material realizations, where the effective frequency is shifted and the level structure is reorganized \citep{Belouad2016GateTunable}.
By contrast, non-Abelian gauge fields introduce an intrinsically matrix-valued structure.
Indeed, they are central to Yang--Mills theory \citep{Yang1954ConservationIsotopic,Ryder1996QuantumField} and also arise as effective gauge potentials in engineered quantum systems, including ultracold atoms and spin--orbit-coupled condensed-matter platforms \citep{Dalibard2011NeutralAtoms,Goldman2014FieldsUltracold}.
This perspective is equally relevant for condensed-matter realizations of Dirac Hamiltonians.
In particular, there is a useful correspondence between the planar Dirac oscillator and gapped graphene, because the same confinement substitution that defines the oscillator can be written in the effective graphene Hamiltonian once the relativistic mass scale is expressed through the gap parameter \citep{Belouad2016GateTunable,Novoselov2005TwoDimensional,Castroneto2009ElectronicProperties}.
Bilayer graphene goes one step further: spatially modulated interlayer couplings generate effective non-Abelian gauge potentials in the low-energy theory \citep{Sanjose2012GrapheneBilayers}.
Consequently, graphene-based systems provide a natural setting in which the Abelian Dirac-oscillator correspondence can be extended to a gauge-field implementation with matrix-valued internal structure.
Although the present model is not intended as a microscopic description of a specific sample, it nevertheless offers an analytical framework for discussing how such backgrounds split internal multiplets, reorganize level structures, and reduce Abelian degeneracies.
Against this background, extending the Dirac oscillator to an external $\mathrm{SU}(2)$ field is both formally natural and physically motivated.
The gauge-theoretic interpretation of relativistic wave equations is itself well established.
In particular, the Dirac equation acquires local gauge covariance through the replacement $\partial_{\mu}\to D_{\mu}$, while non-Abelian generalizations introduce internal generators and path-dependent phase factors that encode parallel transport in internal space \citep{Ryder1996QuantumField,Wu1975NonintegrableFields}.
Moreover, closely related formulations have also been used to describe spin dynamics and effective $U(1)\times SU(2)$ structures in nonrelativistic systems \citep{Frohlich1993InvarianceCurrent}.
In the present context, these ideas clarify why the non-Abelian Dirac oscillator must be formulated on a tensor-product Hilbert space and why the commutator term in the field strength carries direct spectral consequences.
Our objective, therefore, is to identify which spectral properties of the Abelian Dirac oscillator survive under a controlled non-Abelian extension and which are qualitatively modified by the matrix structure of the background.
To this end, we make three points explicit from the outset.
First, spacetime, spatial, and internal indices are kept distinct throughout.
Second, the matter field is treated as a Dirac spinor carrying an additional isospin index, rather than as an ordinary four-component spinor.
Third, the Abelian benchmark is obtained by projecting the full connection onto its commuting $U(1)$ sector, not by setting all couplings in the complete theory to zero.

With these conventions in place, we derive the non-Abelian Hamiltonian, identify the commutator-generated contribution to $\mathcal{F}_{\mu\nu}$, and show how it induces an internal-Zeeman splitting when the background selects a preferred generator in the Lie algebra.
We then construct an aligned planar background for which the spectrum can be obtained analytically.
Far from being merely illustrative, this example provides a benchmark problem that isolates the genuinely non-Abelian contribution and clarifies how Abelian degeneracies are replicated, split, or mixed in the presence of internal degrees of freedom.
Accordingly, the central analytical result of the paper is the exact aligned-planar spectrum
\(
E_{\pm,n,\ell,m_t}=\pm E^{(0)}_{n,\ell}-\zeta m_t,
\)
which serves as the main spectral benchmark for the rest of the discussion.

Unless stated otherwise, the exact closed-form results established in this work refer to this aligned benchmark background, whereas more general $SU(2)$ configurations are discussed at the structural and perturbative levels.

The paper is organized as follows.
Section~II reviews the Dirac oscillator, fixes the notation, and introduces the gauge structure.
Section~III derives the non-Abelian Dirac-oscillator Hamiltonian, discusses the Abelian benchmark, presents an explicit solvable planar background, and supplements the exact formula with a spectral visualization as a function of $\zeta$.
Section~IV then reformulates the main result as a graphene correspondence: monolayer graphene reproduces the Abelian Dirac-oscillator structure, while bilayer graphene provides a natural effective implementation of the non-Abelian gauge-field extension.
Finally, Section~V summarizes the main results and limitations of the present analysis.

Unless stated otherwise, we work in natural units with $\hbar=c=1$ and adopt the metric signature $\eta_{\mu\nu}=\mathrm{diag}(1,-1,-1,-1)$.
In addition, to avoid ambiguity, we fix the notation at the beginning. Greek indices $\mu,\nu,\rho,\sigma=0,1,2,3$ denote spacetime components.
Latin indices $i,j,k=1,2,3$ denote spatial components in $(3+1)$ dimensions, whereas in the planar $(2+1)$-dimensional realization they are restricted to $i,j=1,2$.
Latin indices $a,b,c=1,2,3$ label internal $SU(2)$ directions in the adjoint algebra.
The non-Abelian matter field is written as $\Psi_{\alpha A}(x)$, where $\alpha=1,\ldots,4$ is the Dirac-spinor index and $A=1,2$ is the isospin index in the fundamental representation.
Equivalently, $\Psi(x)\in\mathbb{C}^{4}\otimes\mathbb{C}^{2}$. The Dirac matrices $\gamma^{\mu}$ act on the spinor index only, whereas the generators $T^{a}=\tau^{a}/2$ act on the isospin index only.

\section{The Abelian Dirac oscillator}

The free Dirac equation reads \citep{Ryder1996QuantumField}
\begin{equation}
\left(i\gamma^{\mu}\partial_{\mu}-m\right)\psi=0,
\label{eq:dirac_free}
\end{equation}
where $\gamma^{\mu}$ are the Dirac matrices, $\partial_{\mu}$ is the four-gradient operator, and $m$ is the mass of the particle.
These matrices satisfy the Clifford-algebra relations
\begin{equation}
\{\gamma^{\mu},\gamma^{\nu}\}=2\eta^{\mu\nu}I_{4},
\label{eq:clifford}
\end{equation}
and the antisymmetric tensor $\sigma^{\mu\nu}$ is defined by
\begin{equation}
\sigma^{\mu\nu}=\frac{i}{2}[\gamma^{\mu},\gamma^{\nu}].
\label{eq:sigma}
\end{equation}
In particular, $\sigma^{0i}=i\alpha^{i}$, where $\alpha^{i}=\gamma^{0}\gamma^{i}$.
These structures underlie both the Abelian Dirac oscillator and the non-Abelian extension developed below.

The standard Dirac oscillator is introduced through the non-minimal substitution $\mathbf{p}\rightarrow\mathbf{p}-im\omega\beta\mathbf{r}$, which yields the Hamiltonian
\begin{equation}
H_{\mathrm{DO}}=\boldsymbol{\alpha}\cdot\left(\mathbf{p}-im\omega\beta\mathbf{r}\right)+\beta m.
\label{eq:H_DO}
\end{equation}
Moshinsky and Szczepaniak originally proposed this model as an exactly solvable relativistic bound-state system \citep{Moshinsky1989DiracOscillator}.
Closely related precursor and follow-up analyses include the linear relativistic model of Ito \emph{et al.} and later discussions of exact and supersymmetric aspects of the Dirac oscillator \citep{Ito1967ExampleDynamical,Benitez1990DiracOscillator,Jagannathan1990DiracOscillator}.
In covariant language, the same oscillator term may be re-expressed as a Pauli-type interaction with an effective linearly growing background field.
Consequently, the substitution above and the interaction $\sigma^{\mu\nu}F_{\mu\nu}$ provide equivalent Abelian representations of the same dynamics.

Finally, the complete relativistic spectrum can be written in terms of the radial quantum number $n=0,1,2,\ldots$, the orbital angular momentum $l$, and the total angular momentum $j=l\pm\tfrac{1}{2}$.
Defining $N=2n+l$, one finds \citep{Moshinsky1989DiracOscillator}
\begin{equation}
E^{2}_{Njl}=m^{2}+m\omega
\begin{cases}
2(N-j)+1, & l=j-\frac{1}{2},\\[2pt]
2(N+j)+3, & l=j+\frac{1}{2},
\end{cases}
\label{eq:spectrum_abelian}
\end{equation}
with positive- and negative-energy branches $E_{Njl}=\pm\sqrt{E^{2}_{Njl}}$.
Equation~(\ref{eq:spectrum_abelian}) provides the Abelian benchmark that any consistent non-Abelian extension must reproduce after projection onto the commuting $U(1)$ sector.

\section{FORMULATION OF THE NON-ABELIAN DIRAC OSCILLATOR}

We now introduce the gauge structure with the isospin index made explicit from the beginning.
The total background connection is written as
\begin{equation}
\mathcal{A}_{\mu}(x)=eA^{(0)}_{\mu}(x)I_{2}+gA^{a}_{\mu}(x)T^{a},
\label{eq:conn}
\end{equation}
where $A^{(0)}_{\mu}$ is the Abelian $U(1)$ potential, $A^{a}_{\mu}$ is the $SU(2)$ background, $e$ is the Abelian coupling, $g$ is the non-Abelian coupling, and $T^{a}=\tau^{a}/2$ satisfy $[T^{a},T^{b}]=i\epsilon^{abc}T^{c}$.
The covariant derivative acting on $\Psi_{\alpha A}$ is therefore
\begin{equation}
\mathcal{D}_{\mu}=\partial_{\mu}+i\mathcal{A}_{\mu}=\partial_{\mu}+ieA^{(0)}_{\mu}I_{2}+igA^{a}_{\mu}T^{a}.
\label{eq:cov_derivative}
\end{equation}
The corresponding field strength is defined by
\begin{equation}
\mathcal{F}_{\mu\nu}\equiv \frac{1}{i}[\mathcal{D}_{\mu},\mathcal{D}_{\nu}]=eF^{(0)}_{\mu\nu}I_{2}+gF^{a}_{\mu\nu}T^{a},
\label{eq:field_strength}
\end{equation}
with
\begin{equation}
F^{(0)}_{\mu\nu}=\partial_{\mu}A^{(0)}_{\nu}-\partial_{\nu}A^{(0)}_{\mu},
\label{eq:field_strength_u1}
\end{equation}
\begin{equation}
F^{a}_{\mu\nu}=\partial_{\mu}A^{a}_{\nu}-\partial_{\nu}A^{a}_{\mu}-g\epsilon^{abc}A^{b}_{\mu}A^{c}_{\nu}.
\label{eq:field_strength_su2}
\end{equation}
Equation~(\ref{eq:field_strength_su2}) makes the commutator contribution explicit. This term vanishes in the Abelian sector but survives for spatially uniform matrix potentials whenever the internal components fail to commute.
As a result, non-Abelian backgrounds can generate physical effects even when the derivative part of the field strength is absent.
Related relativistic and Pauli-type formulations for particles carrying non-Abelian charge have been developed in complementary settings \citep{Dossa2020ParticleCarrying,Dossa2020PauliHamiltonian}, while condensed-matter analogues of matrix-valued gauge connections have been discussed from Yang--Mills and spin-transport viewpoints \citep{Berche2012ClassicalYang,Tan2011SpinFields,Tan2020YangMills}.
The minimally coupled Dirac equation on $\mathbb{C}^{4}\otimes\mathbb{C}^{2}$ takes the form
\begin{equation}
\left[i\gamma^{\mu}\mathcal{D}_{\mu}-m\right]\Psi=0.
\label{eq:dirac_minimal_na}
\end{equation}
To recover the covariant oscillator construction, we supplement it with the Pauli-type interaction used in the standard derivation of the Dirac oscillator:
\begin{equation}
\left[i\gamma^{\mu}\mathcal{D}_{\mu}+\frac{\kappa}{4m}\sigma^{\mu\nu}\mathcal{F}_{\mu\nu}-m\right]\Psi=0.
\label{eq:dirac_pauli_na}
\end{equation}
Here $\sigma^{\mu\nu}$ acts on the Dirac-spinor index, whereas $\mathcal{F}_{\mu\nu}$ acts on the isospin index.
This tensor-product structure is essential and will be preserved throughout the Hamiltonian formulation.

For the explicit background considered below, we separate the Abelian oscillator-generating contribution from a static matrix-valued $SU(2)$ part.
The Abelian piece is chosen as
\begin{equation}
A^{(0)}_{\mu}(x)=\frac{\lambda}{2}\left[(u\!\cdot\!x)x_{\mu}-\frac{x^{2}}{2}u_{\mu}\right],\qquad u^{\mu}=(1,0,0,0),
\label{eq:ab_potential}
\end{equation}
for which $F^{(0)}_{0i}=\lambda x_{i}$ and $F^{(0)}_{ij}=0$.
The non-Abelian background is parameterized by
\begin{equation}
A^{a}_{0}=\lambda\phi^{a},\qquad A^{a}_{i}=\eta\,\Xi_{i}^{\ a},
\label{eq:na_background}
\end{equation}
where $\phi^{a}$ and $\Xi_{i}^{\ a}$ are dimensionless constants specifying the orientation of the $SU(2)$ background.
For this choice, the non-Abelian field-strength components become
\begin{equation}
gF^{a}_{0i}T^{a}=-g^{2}\lambda\eta\,\epsilon^{abc}\phi^{b}\Xi_{i}^{\ c}T^{a},
\label{eq:na_0i}
\end{equation}
\begin{equation}
gF^{a}_{ij}T^{a}=-g^{2}\eta^{2}\,\epsilon^{abc}\Xi_{i}^{\ b}\Xi_{j}^{\ c}T^{a}.
\label{eq:na_ij}
\end{equation}
This choice is deliberately minimal.
On the Abelian side, it reproduces the electric background that generates the Dirac oscillator.
On the non-Abelian side, it isolates the commutator contribution by suppressing derivative-driven inhomogeneities.
Consequently, the spectrum obtained below answers a precise question: what changes when the oscillator is coupled to a field strength whose nontrivial content is generated directly by the internal algebra?

Starting from Eq.~(\ref{eq:dirac_pauli_na}), multiplying by $\gamma^{0}=\beta$, and using $\sigma^{0i}=i\alpha^{i}$, we obtain the Hamiltonian equation
\begin{equation}
i\partial_{t}\Psi=\hat{H}\Psi.
\end{equation}
Because $\Psi$ carries both Dirac and isospin indices, the Hamiltonian acts on both sectors.
Writing tensor products explicitly, one finds
\begin{equation}
\hat{H}=H_{\mathrm{DO}}\otimes I_{2}+\hat{V}_{\mathrm{NA}},
\label{eq:H_NA_full}
\end{equation}
where $H_{\mathrm{DO}}$ is the Abelian Dirac-oscillator Hamiltonian in Eq.~(\ref{eq:H_DO}) and
\begin{equation}
\hat{V}_{\mathrm{NA}}=\frac{\kappa}{4m}\,\beta\,\sigma^{\mu\nu}\,gF^{a}_{\mu\nu}T^{a}.
\label{eq:V_NA}
\end{equation}
For the background defined in Eqs.~(\ref{eq:na_0i})--(\ref{eq:na_ij}), this becomes
\begin{equation}
\hat{V}_{\mathrm{NA}}=-i\frac{\kappa g^{2}\lambda\eta}{2m}\,\beta\alpha^{i}\epsilon^{abc}\phi^{b}\Xi_{i}^{\ c}T^{a}-\frac{\kappa g^{2}\eta^{2}}{4m}\,\beta\sigma^{ij}\epsilon^{abc}\Xi_{i}^{\ b}\Xi_{j}^{\ c}T^{a}.
\label{eq:V_NA_explicit}
\end{equation}
For $0<\eta\ll1$, one also has $\eta^{2}\ll\eta\ll1$, because multiplying the inequality $0<\eta\ll1$ by the additional small factor $\eta$ suppresses the quadratic term even further. Accordingly, the first term is linear in $\eta$, whereas the second term is quadratic in $\eta$ and originates directly from the commutator structure of the non-Abelian field strength.
This hierarchy is important: the linear contribution acts primarily as a channel-dependent displacement of the non-minimal coupling, while the quadratic term is the first one capable of producing a genuine internal-Zeeman splitting.

Using the Abelian identification $\lambda=2m^{2}\omega/(\kappa e)$, the weak-background Hamiltonian may be written as
\begin{equation}
\hat{H}\simeq H_{\mathrm{DO}}\otimes I_{2}-im\omega\,\beta\alpha^{i}\,\Gamma_{i},\qquad \Gamma_{i}\equiv\frac{g^{2}\eta}{e}\,\epsilon^{abc}\phi^{b}\Xi_{i}^{\ c}T^{a},
\label{eq:H_NA_linear}
\end{equation}
or, in component form,
\begin{equation}
i\partial_{t}\Psi_{\alpha A}=\left[(H_{\mathrm{DO}})_{\alpha}{}^{\beta}\,\delta_{A}{}^{B}-im\omega\,(\beta\alpha^{i})_{\alpha}{}^{\beta}(\Gamma_{i})_{A}{}^{B}\right]\Psi_{\beta B}.
\label{eq:component_form}
\end{equation}
Equation~(\ref{eq:component_form}) is the corrected non-Abelian Dirac-oscillator equation: the ordinary Dirac oscillator acts on the spinor sector, whereas the matrix $\Gamma_{i}$ acts on the isospin sector.

The same tensor-product structure remains valid in the planar $(2+1)$-dimensional realization.
One simply restricts the spatial indices to $i,j=1,2$ and sets $\mathbf{r}=(x,y)$, while the field still carries both a Dirac index and an isospin index.
Therefore, spatial labels and internal labels remain conceptually distinct even in the reduced geometry.

To render the spectral discussion fully explicit, we consider the simplest aligned planar background in $(2+1)$ dimensions for which the commutator term is nonzero while the algebra remains tractable.
We choose
\begin{equation}
A^{a}_{0}=0,\qquad A^{1}_{1}=\eta,\qquad A^{2}_{2}=\eta,\qquad \text{all other }A^{a}_{\mu}=0.
\label{eq:aligned_background}
\end{equation}
For this background, the only nonvanishing non-Abelian field-strength component is
\begin{equation}
F^{3}_{12}=-g\eta^{2},\qquad F^{a}_{0i}=0,\qquad F^{a}_{12}=0\ \text{for}\ a\neq3.
\label{eq:aligned_fieldstrength}
\end{equation}
Hence the Pauli term reduces to a single internal operator proportional to $T_{3}$.
In the planar Dirac algebra, one has $\beta\sigma^{12}=I_{2}$ up to the sign convention associated with the two inequivalent irreducible representations, so the Hamiltonian becomes
\begin{equation}
\hat H^{(2+1)}_{\rm aligned}=H^{(2+1)}_{\rm DO}\otimes I_{2}-\zeta\,I_{\rm D}\otimes T_{3},\qquad \zeta\equiv \frac{\kappa g^{2}\eta^{2}}{2m}.
\label{eq:H_aligned}
\end{equation}
Since $T_{3}\chi_{m_t}=m_t\chi_{m_t}$ with $m_t=\pm\tfrac12$ in the fundamental representation, the problem factorizes into two independent isospin channels.
If $E^{(0)}_{n,\ell}$ denotes an exact planar Dirac-oscillator level, then the aligned non-Abelian background yields the closed-form spectrum
\begin{equation}
E_{\pm,n,\ell,m_t}=\pm E^{(0)}_{n,\ell}-\zeta m_t,\qquad m_t=\pm\frac12.
\label{eq:explicit_splitting}
\end{equation}
Therefore, the two isospin components of each planar Abelian level are separated by
\begin{equation}
\Delta E_{n,\ell}=E_{\pm,n,\ell,+1/2}-E_{\pm,n,\ell,-1/2}=-\zeta,
\label{eq:explicit_gap}
\end{equation}
with magnitude $|\Delta E_{n,\ell}|=\zeta$.
This result shows explicitly that the first genuinely non-Abelian splitting is quadratic in the background amplitude and is controlled by the commutator term in Eq.~(\ref{eq:field_strength_su2}).
It is also the principal exact spectral benchmark of the present work: the aligned background preserves the Abelian ordering through $E^{(0)}_{n,\ell}$ while resolving each level into internal channels with universal linear slopes in $\zeta$.

The same argument extends immediately to higher internal representations.
For isospin $t$, the aligned background generates $2t+1$ branches labeled by $m_t=-t,-t+1,\ldots,t$, each shifted by $-\zeta m_t$.
Thus the aligned planar solution provides both an exact spectrum and a representation-theoretic rule for how Abelian levels are replicated and reorganized when the internal algebra becomes dynamical.

At the same time, this exact solvability should not be overextended: the closed-form expression above is established for the aligned benchmark background, not for arbitrary non-Abelian configurations.

The aligned configuration should be viewed as the simplest uniform benchmark.
Nevertheless, a natural extension, which we leave for future work, is to investigate whether suitably chosen \emph{non-uniform} non-Abelian field profiles can also sustain Landau-like level structures.
This question is motivated by recent Abelian results for Dirac electrons in graphene, where explicit non-uniform magnetic fields were shown to remain isospectral to the Landau problem \citep{Ghosh2025LandauLevels}.

\paragraph{Abelian benchmark and expected non-Abelian effects.}
The Abelian benchmark is obtained by projecting the full connection onto its commuting $U(1)$ component,
\begin{equation}
\mathcal{A}_{\mu}\longrightarrow eA^{(0)}_{\mu}I_{2},\qquad A^{a}_{\mu}\longrightarrow0.
\label{eq:abelian_projection}
\end{equation}
Under this restriction,
\begin{equation}
\mathcal{F}_{\mu\nu}\longrightarrow eF^{(0)}_{\mu\nu}I_{2},\qquad \hat{H}\longrightarrow H_{\mathrm{DO}}\otimes I_{2}.
\label{eq:abelian_H}
\end{equation}
Hence the internal multiplet does not disappear.
Instead, it decouples into identical copies of the Abelian Dirac oscillator.
In the fundamental representation, the projected theory therefore carries a trivial twofold isospin multiplicity on top of the usual Dirac-oscillator spectrum.
By contrast, setting both $e$ and $g$ to zero in the complete theory would return the free Dirac equation, not the Abelian Dirac-oscillator limit.

For the Abelian Dirac oscillator, the exact spectrum is given by Eq.~(\ref{eq:spectrum_abelian}).
A convenient basis in the projected sector is $\Psi^{(0)}_{\alpha A}(x)=\psi^{(0)}_{\alpha}(x)\chi_{A}$, where $\chi_{A}$ is a constant isospinor.
Once the $SU(2)$ background is turned on, the matrix operators in Eqs.~(\ref{eq:H_NA_linear}) and (\ref{eq:H_aligned}) either split or mix this otherwise trivial internal multiplicity. The aligned planar result in Eq.~(\ref{eq:explicit_splitting}) shows the mechanism in its simplest form: the projected Abelian spectrum is copied into two isospin channels and then shifted by opposite amounts according to the eigenvalue of $T_{3}$.

The decisive qualitative difference between the Abelian and non-Abelian theories originates from the commutator term in the field strength.
Even for spatially uniform matrix potentials, $[A_{\mu},A_{\nu}]\neq0$ can generate a nonvanishing field strength and therefore observable consequences.
When the background is aligned with a single generator, the non-Abelian sector acts as an effective Zeeman field in internal space.
Consequently, the problem decomposes into channels labeled by the eigenvalues of that generator, and the Abelian levels split into internal multiplets.
\begin{figure}[t]
\centering
\IfFileExists{Figure_spectrum.png}{\includegraphics[width=0.88\linewidth]{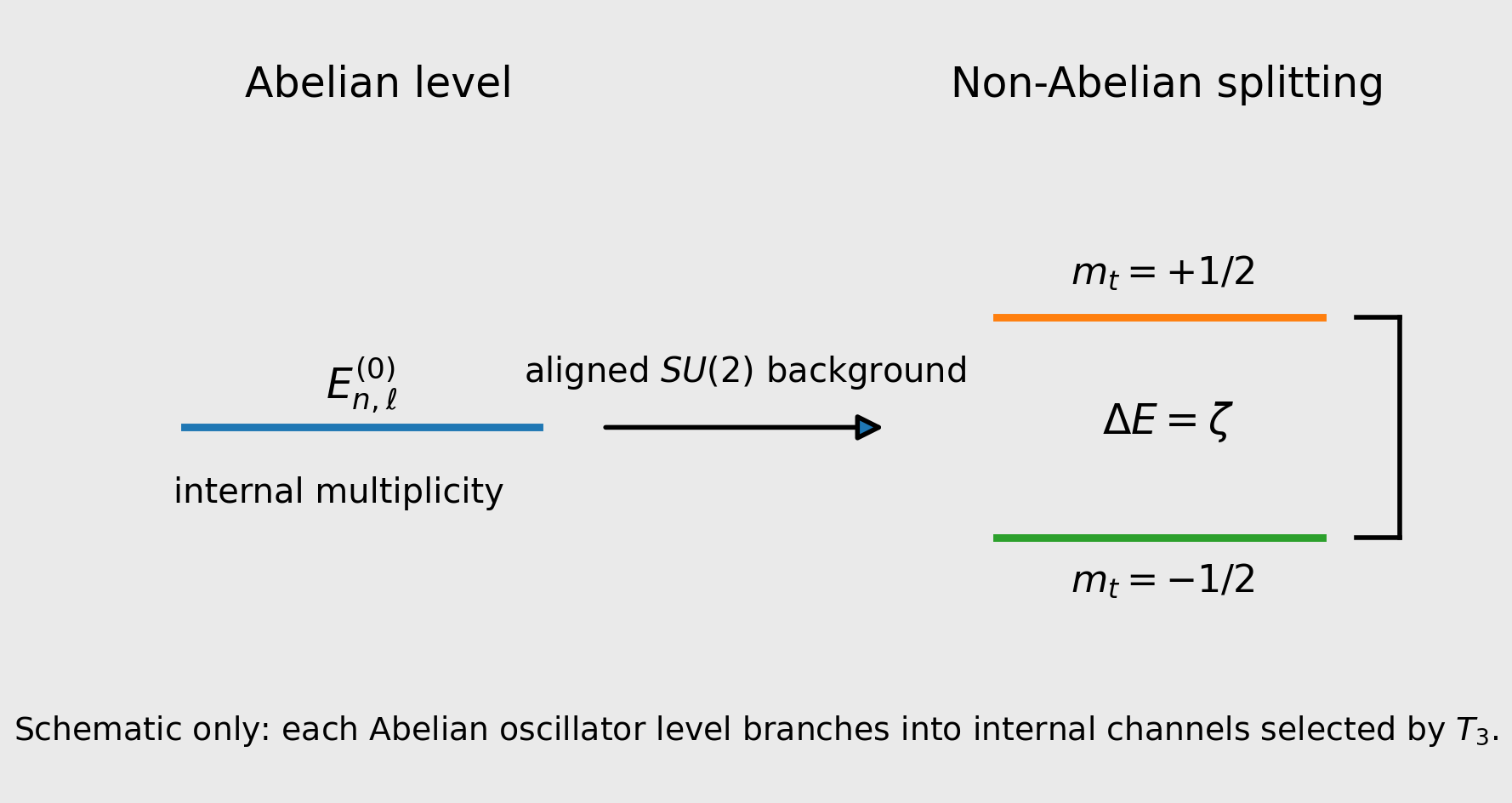}}{\fbox{\parbox{0.82\linewidth}{\centering Schematic spectrum figure placeholder.\\The source image Figure\_spectrum.png was not present in the uploaded files, so a placeholder is shown for compilation.}}}
\caption{Schematic representation of the internal-Zeeman mechanism in the non-Abelian Dirac oscillator.
The left side shows a single Abelian Dirac-oscillator level $E_{Njl}$, which serves as the organizing spectral backbone of the theory.
The right side shows the same level after coupling to an aligned $\mathrm{SU}(2)$ background that selects the generator $T_{3}=\tau_{3}/2$.
In the fundamental representation, the internal degeneracy is resolved into two branches labeled by $m_{t}=\pm\tfrac{1}{2}$.
Thus the figure visualizes the transition from a purely orbital--spin classification to a combined orbital--spin--isospin classification.
It also clarifies the role of the explicit planar solution: Eq.~(\ref{eq:explicit_splitting}) provides the analytic realization of the schematic splitting displayed here.}
\label{fig:internal_zeeman_split}
\end{figure}

Figure~\ref{fig:internal_zeeman_split} plays a structural role in the manuscript rather than merely an illustrative one.
It summarizes, in a single diagram, the logic developed across the formalism.
First, it identifies the Abelian Dirac-oscillator level as the reference quantity that survives in the projected $U(1)$ sector.
Second, it shows how the non-Abelian background does not destroy that spectral backbone, but reorganizes it into internal branches.
Third, it anticipates the exact planar result by making clear that the splitting is controlled by the eigenvalues of the selected generator.
Accordingly, the figure provides the conceptual bridge between the general Hamiltonian construction, the perturbative discussion of weak backgrounds, and the exact solution obtained for the aligned planar configuration.

\begin{figure}[t]
\centering
\IfFileExists{Figure_spectrum_vs_zeta_reproduced.png}{\includegraphics[width=0.82\linewidth]{Figure_spectrum_vs_zeta_reproduced.png}}{\fbox{\parbox{0.78\linewidth}{\centering Spectrum-vs-zeta figure placeholder.\\The file Figure\_spectrum\_vs\_zeta.png was not present in the working directory, so a placeholder is shown for compilation.}}}
\caption{Representative visualization of the exact aligned-planar spectrum as a function of the non-Abelian splitting parameter $\zeta$ for four sample Abelian backbone levels $(n,\ell)=(0,0),(0,1),(1,0),(1,1)$ in normalized units.
For each fixed $(n,\ell)$, the positive- and negative-energy sectors split into two isospin branches $m_t=\pm\tfrac12$ according to the exact rule
$E_{\pm,n,\ell,m_t}=\pm E^{(0)}_{n,\ell}-\zeta m_t$.
The plot therefore makes two properties of Eq.~(\ref{eq:explicit_splitting}) explicit: the splitting width within each internal doublet is $|\Delta E|=\zeta$, while the dependence on $(n,\ell)$ enters only through the Abelian intercepts $\pm E^{(0)}_{n,\ell}$.
The displayed backbone values are representative and are used only to visualize the universal linear branching pattern.}
\label{fig:spectrum_vs_zeta}
\end{figure}

Figure~\ref{fig:spectrum_vs_zeta} complements the schematic level diagram in Fig.~\ref{fig:internal_zeeman_split} with a direct rendering of the exact formula.
Accordingly, it should be read as a visualization of Eq.~(\ref{eq:explicit_splitting}), not as an independent approximation scheme.
Its message is simple but important: in the aligned benchmark, the non-Abelian background does not distort the Abelian level ordering.
Instead, it duplicates each positive- and negative-energy Abelian level into two parallel internal channels, with universal slopes fixed by $m_t=\pm\tfrac12$ and a separation that grows linearly with $\zeta$.
For that reason, the aligned solution provides a particularly useful reference problem for the broader theory.

This interpretation is particularly transparent at the level of the squared Hamiltonian.
The Pauli-type coupling $\sigma^{\mu\nu}\mathcal{F}_{\mu\nu}$ contains operators proportional to the internal generator selected by the background.
Therefore, the non-Abelian sector acts as an effective Zeeman field in isospin space.
In the fundamental representation one has $T_{3}=\tau_{3}/2$ with eigenvalues $\pm\tfrac{1}{2}$, and each Abelian level $E_{Njl}$ branches into two components.
More generally, an isospin-$t$ multiplet splits into $2t+1$ branches. The overall size of the splitting is governed by the internal part of the field strength and is accompanied by a partial lifting of the Abelian degeneracies whenever the background breaks the accidental symmetries of the original oscillator.
For weak non-Abelian backgrounds, the leading shifts can be estimated perturbatively.
Let $\ket{\Psi^{(0)}}$ be an eigenstate of the Abelian Hamiltonian $H_{\mathrm{DO}}$ with energy $E^{(0)}$.
Writing the full Hamiltonian as $H$ and expanding $H^{2}=H_{\mathrm{DO}}^{2}+\delta(H^{2})$, the first-order energy correction is
\begin{equation}
\Delta E\simeq\frac{\langle\Psi^{(0)}|H^{2}-H_{\mathrm{DO}}^{2}|\Psi^{(0)}\rangle}{2E^{(0)}},
\label{eq:deltaE}
\end{equation}
which identifies which degeneracies are lifted by a given $SU(2)$ background.
Contributions diagonal in $T_{3}$ split multiplets, whereas off-diagonal components mix internal states and can generate avoided crossings when external parameters are varied.
In the weak-coupling regime, one may use unperturbed tensor-product states
\begin{equation}
\left|\Psi^{(0)}\right\rangle=\left|Njlm_{j}\right\rangle\otimes\left|tm_{t}\right\rangle,
\end{equation}
and evaluate the leading corrections from
\begin{equation}
\Delta(E^{2})\simeq\left\langle \Psi^{(0)}\right|\left(H^{2}-H_{\mathrm{DO}}^{2}\right)\left|\Psi^{(0)}\right\rangle,
\qquad
\Delta E\simeq\frac{\Delta(E^{2})}{2E^{(0)}},
\label{eq:pert_shift_E2}
\end{equation}
where $E^{(0)}$ is the corresponding Abelian energy.
In practice, within each degenerate Abelian level one diagonalizes the non-Abelian correction in internal space.
This procedure makes explicit which degeneracies are symmetry-protected and which are lifted by the background.
\paragraph{Symmetry and angular-momentum classification.}
For the Abelian Dirac oscillator (i.e., $\eta=0$), the total angular momentum
\[
\hat{\mathbf{J}}=\hat{\mathbf{L}}+\frac{\hat{\boldsymbol{\Sigma}}}{2}
\]
is conserved, and the corresponding energy spectrum may therefore be classified by the quantum number $j$. In particular, one finds
\begin{align}
[\hat{\mathbf{L}},H_{\mathrm{DO}}]
&= i\,(\boldsymbol{\alpha}\times\hat{\mathbf{p}})
   - m\,(\hat{\mathbf{r}}\times\boldsymbol{\alpha})\,\beta, \\
\left[\frac{\hat{\boldsymbol{\Sigma}}}{2},H_{\mathrm{DO}}\right]
&= -i\,(\boldsymbol{\alpha}\times\hat{\mathbf{p}})
   + m\,(\hat{\mathbf{r}}\times\boldsymbol{\alpha})\,\beta,
\end{align}
so that
\[
[\hat{\mathbf{J}},H_{\mathrm{DO}}]=0.
\]
Here, $\boldsymbol{\Sigma}$ is related to $\boldsymbol{\alpha}$ through
\[
\boldsymbol{\Sigma}=-\frac{i}{2}\,\boldsymbol{\alpha}\times\boldsymbol{\alpha}.
\]

For $\eta\neq0$, the status of $\hat{\mathbf{J}}$ depends on the spatial structure of the non-Abelian background encoded in $\Xi_{i}{}^{a}$. A generic fixed tensor $\Xi_{i}{}^{a}$ selects preferred directions and may therefore break ordinary rotational invariance, so that $j$ ceases to be a good quantum number. By contrast, when the background is invariant under simultaneous spatial and internal rotations, it is natural to introduce the combined (grand) angular momentum
\begin{equation}
\hat{\mathbf{K}}=\hat{\mathbf{L}}+\frac{\hat{\boldsymbol{\Sigma}}}{2}+\hat{\mathbf{T}},
\label{eq:grand_spin}
\end{equation}
where $\hat{\mathbf{T}}$ denotes the generators of $SU(2)$ acting on the internal multiplet. In such symmetric configurations, the spectrum organizes into multiplets labeled by $K$, and the comparison with the Abelian Dirac oscillator becomes especially transparent: the non-Abelian interaction produces controlled splittings and rearrangements of the Abelian levels, determined by the internal representation and by the symmetry of the chosen gauge background.
\section{GRAPHENE CORRESPONDENCE AND NON-ABELIAN IMPLEMENTATION}
The purpose of this section is to make explicit a two-step correspondence.
First, at the Abelian level, the planar Dirac oscillator and gapped graphene share the same effective Dirac structure once the oscillator mass parameter is re-expressed in terms of the graphene gap scale.
Second, when the internal valley or layer doublet is retained, this correspondence extends naturally to the gauge-field case, with matrix-valued effective connections providing a concrete implementation of the non-Abelian model.
In this respect, bilayer graphene is especially useful because its low-energy sector already carries a layer pseudospin on which such connections act \citep{Novoselov2005TwoDimensional,Castroneto2009ElectronicProperties,Sanjose2012GrapheneBilayers}.
In monolayer graphene, the standard continuum description shows that smooth strain primarily enters through valley-opposite pseudogauge fields generated by hopping modulations, so the effective connection is valley diagonal and therefore Abelian at low energy \citep{Castroneto2009ElectronicProperties,Vozmediano2010FieldsGraphene}.
By contrast, topological defects, holonomy effects, intervalley processes, and especially bilayer interlayer couplings can mix internal channels and thereby require genuinely matrix-valued effective gauge potentials \citep{Vozmediano2010FieldsGraphene,Sanjose2012GrapheneBilayers,Ramires2018ElectricallyTunable,Gonzalez2016ConfiningRepulsive}.
This Abelian-versus-non-Abelian distinction is precisely what makes graphene a useful organizing example here: monolayer graphene makes the Dirac-oscillator correspondence transparent, whereas bilayer graphene supplies the more natural arena for its non-Abelian implementation.
For a single valley $\xi=\pm1$, a convenient effective Hamiltonian for gapped graphene in the presence of a real electromagnetic field $\mathbf{A}^{\mathrm{em}}$ and a strain-induced pseudogauge field $\mathbf{A}^{\mathrm{s}}$ is
\begin{equation}
H^{(\xi)}_{\mathrm{G}}=v_{F}\,\boldsymbol{\sigma}_{\xi}\!\cdot\!\left(\mathbf{p}-e\mathbf{A}^{\mathrm{em}}-\xi\mathbf{A}^{\mathrm{s}}\right)+\Delta\,\beta,\qquad \boldsymbol{\sigma}_{\xi}\equiv(\sigma_{x},\xi\sigma_{y}),\qquad \beta\equiv\sigma_{z},
\label{eq:graphene_base}
\end{equation}
where $v_{F}$ is the Fermi velocity and $\Delta$ is a spectral gap, following the standard low-energy gauge-field description of graphene and graphene-like Dirac materials \citep{Castroneto2009ElectronicProperties,Vozmediano2010FieldsGraphene}.
In direct analogy with the planar Dirac oscillator, one may then introduce the confinement rule
\begin{equation}
\mathbf{p}\longrightarrow \mathbf{p}-i\frac{\Delta\omega}{v_{F}}\,\beta\,\mathbf{r},
\label{eq:graphene_do_sub}
\end{equation}
which yields the effective graphene Dirac-oscillator Hamiltonian
\begin{equation}
H^{(\xi)}_{\mathrm{G,DO}}=v_{F}\,\boldsymbol{\sigma}_{\xi}\!\cdot\!\left(\mathbf{p}-e\mathbf{A}^{\mathrm{em}}-\xi\mathbf{A}^{\mathrm{s}}-i\frac{\Delta\omega}{v_{F}}\,\beta\,\mathbf{r}\right)+\Delta\,\beta.
\label{eq:graphene_do}
\end{equation}
This construction makes the Abelian correspondence explicit: apart from replacing the relativistic mass scale by the graphene gap parameter, the confinement rule is the same as in the planar Dirac oscillator.
In this sense, graphene provides an effective realization of Dirac-oscillator kinematics and confined spectra, consistent with earlier graphene quantum-dot analyses \citep{Belouad2016GateTunable}.
We stress, however, that the discussion in this section is not intended as a microscopic derivation for a specific graphene device.
Rather, it should be read as an effective-theory correspondence: the Abelian confinement rule reproduces the planar Dirac-oscillator structure, while matrix-valued valley or layer connections motivate the non-Abelian extension at the low-energy Hamiltonian level.
In the present study, graphene therefore serves primarily as a tractable Dirac system for examining how matrix-valued gauge backgrounds reorganize confined Dirac spectra, rather than as a fully microscopic material-specific model.

Within this effective-theory perspective, the gauge-field extension of the correspondence follows by allowing the effective gauge potential to act on a valley or layer doublet.
Writing the low-energy doublet as $\Upsilon$ and the matrix-valued connection as $\mathcal{A}^{\mathrm{NA}}_{i}=A^{a}_{i}T^{a}$, one obtains the schematic realization
\begin{equation}
\hat{H}_{\mathrm{G,NA}}=H_{\mathrm{G,DO}}\otimes I_{2}+v_{F}\,\sigma_{i}\otimes A^{a}_{i}T^{a}+\hat{V}_{\mathrm{comm}},
\label{eq:graphene_na}
\end{equation}
where $\hat{V}_{\mathrm{comm}}$ collects the terms generated by the non-Abelian field strength.
In this form, Eq.~(\ref{eq:graphene_na}) is the graphene-side implementation of the non-Abelian Dirac oscillator: the first term reproduces the Abelian correspondence, the second introduces the matrix-valued connection, and $\hat{V}_{\mathrm{comm}}$ isolates the genuinely non-Abelian commutator contribution.
In bilayer graphene this internal sector is particularly natural because the relevant low-energy states already form a layer doublet, and spatially modulated interlayer couplings provide an explicit route to effective non-Abelian gauge potentials \citep{Sanjose2012GrapheneBilayers,Ramires2018ElectricallyTunable,Gonzalez2016ConfiningRepulsive}.
Consequently, the aligned splitting mechanism derived for the abstract $SU(2)$ model acquires a direct interpretation as a controlled lifting of valley or layer degeneracy by a commutator-generated internal field.
If the effective background selects one generator, for example $T_{3}$, the confined graphene or bilayer-graphene levels separate into internal branches according to the same rule as in the aligned planar model,
\begin{equation}
E^{\mathrm{G}}_{n,\xi,m_{t}}\approx E^{(0)}_{n,\xi}-\zeta m_{t},
\label{eq:graphene_split}
\end{equation}
where $E^{(0)}_{n,\xi}$ denotes the corresponding Dirac-material level before valley or layer mixing is switched on.
The most direct signatures are therefore valley- or layer-resolved splittings, possible avoided crossings when off-diagonal internal terms are present, and channel-dependent phase shifts in coherent transport.

This interpretation should nevertheless be read with care.
The continuum description is valid only near the Dirac points, and smooth strain by itself usually produces an effectively Abelian pseudogauge field rather than a genuinely non-Abelian one \citep{Castroneto2009ElectronicProperties,Vozmediano2010FieldsGraphene}.
Accordingly, the full commutator-driven sector of the present theory is best viewed as an effective implementation that becomes relevant only when an intervalley or interlayer mechanism makes different internal components of the connection fail to commute.
Furthermore, for these graphene-based implementations, the parameter playing the role of the oscillator mass is often best interpreted through an effective, energy-dependent Dirac-material mass scale rather than through a literal bare band mass \citep{Ariel2012EffectiveMassEnergyMass,ArielNatan2012ElectronEffectiveMassGraphene,Jiang2007InfraredLandau,Boumali2015ThermodynamicGraphene}.

Beyond graphene, the planar setting also connects naturally with Jaynes--Cummings (JC) and anti-Jaynes--Cummings (AJC) mappings, magnetic-field Dirac-oscillator realizations, and related analogue simulators.
The Dirac-oscillator framework preserves the algebraic structure of the JC and AJC models \citep{JaynesCummings1963BeamMaser,ShoreKnight1993JaynesCummings}, thereby making contact with well-established quantum-optical realizations and analog-simulation paradigms.
In particular, Berm\'udez \emph{et al.} \citep{Bermudez2007MappingDirac,Bermudez2008ChiralityTransition} obtained an exact solution of the $(2+1)$-dimensional Dirac oscillator and established a precise mapping to a JC model describing the interaction of a two-level system with a single quantized mode.
They also proposed a trapped-ion realization of the 2D Dirac oscillator, making relevant observables accessible within contemporary quantum-simulation capabilities.
Complementary realizations of related Dirac dynamics have been achieved in microwave-resonator arrays and trapped-ion systems \citep{Francovillafane2013ExperimentalRealization,Gerritsma2010SimulationDirac}, while magnetic-field Dirac-oscillator variants and graphene-based settings provide additional condensed-matter analogues \citep{Belouad2016GateTunable,Castroneto2009ElectronicProperties,Knight2024JOSAB,Mandal2010DiracOscillator,MandalRai2012NoncommutativeExternalMagnetic,Zhang2025ExperimentalSimulationDirac}.
These platforms reinforce the methodological point of the present work: even when a microscopic realization is not literally relativistic, the non-Abelian Dirac oscillator remains a useful benchmark Hamiltonian for organizing commutator-driven level splitting and internal-channel dynamics.

\section{CONCLUSION}
We have developed a covariant formulation of the Dirac oscillator coupled to external non-Abelian gauge fields.
Specifically, by embedding the oscillator interaction in an $\mathrm{SU}(2)$ background, we derived the corresponding non-Abelian field-strength tensor, including the commutator term that encodes the intrinsic nonlinearity of Yang--Mills fields, and obtained the generalized interaction proportional to $\sigma^{\mu\nu}\mathcal{F}_{\mu\nu}$.
As a result, the Hamiltonian acquires additional matrix-valued couplings acting on internal degrees of freedom and therefore extends the familiar Abelian Dirac oscillator in a mathematically consistent way.
Three structural aspects are essential to the revised presentation.
First, the matter field is written explicitly as $\Psi_{\alpha A}$, so the tensor-product structure between Dirac and isospin spaces is visible at every stage of the derivation.
Second, the notation distinguishes systematically between spacetime indices, spatial indices, and internal $SU(2)$ indices.
Third, the Abelian reduction is defined correctly as the projection onto the commuting $U(1)$ background rather than as the trivial vanishing-coupling limit of the full theory.
From the physical side, the main consequence of the non-Abelian extension is the appearance of internal-Zeeman splittings produced by the commutator contribution to the field strength.
This mechanism becomes completely explicit in the aligned planar background, for which the spectrum is obtained in closed form.
In that example, each Abelian Dirac-oscillator level is copied into two isospin channels and shifted by opposite amounts according to the eigenvalue of $T_{3}$.
Accordingly, the exact formula \(E_{\pm,n,\ell,m_t}=\pm E^{(0)}_{n,\ell}-\zeta m_t\) should be regarded as the principal analytical output of the paper, while the additional spectrum-vs-$\zeta$ plot simply renders this result in a form that makes the universal linear branching pattern immediately visible.
Moreover, the streamlined discussion in Section~IV was organized to clarify the condensed-matter meaning of this result: at the Abelian level, the same confinement substitution establishes a correspondence between the planar Dirac oscillator and gapped graphene, while retaining the valley or layer doublet extends that correspondence to the gauge-field case.
In turn, bilayer graphene provides the most natural effective implementation of the non-Abelian Hamiltonian, because its low-energy theory already supports matrix-valued couplings between internal channels.
At the same time, the graphene discussion should be interpreted with the appropriate effective-theory scope: the manuscript does not claim a full microscopic derivation for a particular sample, but rather an analytically controlled correspondence that clarifies how matrix-valued backgrounds can reorganize confined Dirac spectra.
Nevertheless, the present analysis should be read within a definite scope, since only a specially aligned background is solved exactly.
This restriction is deliberate, because it yields analytical control and cleanly separates linear kinematic displacements from the genuinely non-Abelian quadratic splitting generated by the commutator term.
Within that scope, the model provides a useful benchmark for future studies of less symmetric backgrounds, higher internal representations, curved geometries, time-dependent gauge configurations, and more microscopic graphene or bilayer-graphene realizations of matrix-valued effective gauge fields \citep{Sanjose2012GrapheneBilayers,Vozmediano2010FieldsGraphene}.
In particular, the manuscript should be read as combining an exactly solvable benchmark with a numerical visualization of that exact result, rather than as claiming an independent many-parameter numerical spectroscopy beyond the aligned analytic case.
Therefore, the revised paper advances a definite claim and no broader one: it provides an exact benchmark solution for a specially aligned background, together with a framework for analyzing how more general non-Abelian couplings can split or mix the Abelian Dirac-oscillator spectrum.

\section*{CONFLICTS OF INTEREST STATEMENT}
No conflict of interest was declared by the authors.

\section*{FUNDING STATEMENT}
No funds have been received for this manuscript.

\section*{DATA AVAILABILITY STATEMENT}
In this study, no new data were generated or analyzed.

\bibliographystyle{apsrev4-1}
\bibliography{references_checked_clean}

@misc{Belouad2016GateTunable,
  author  = {Belouad, Abdelhadi and Jellal, Ahmed and Zahidi, Youness},
  title   = {Gate-Tunable Graphene Quantum Dot and Dirac Oscillator},
  journal = {Phys. Lett. A},
  year    = {2016},
  doi     = {10.1016/j.physleta.2015.11.025},
  url     = {https://doi.org/10.1016/j.physleta.2015.11.025},
  note    = {Physics Letters A 380, 773--778 (2016); DOI: 10.1016/j.physleta.2015.11.025}
}

@article{Benitez1990DiracOscillator,
  author  = {Ben{\'i}tez, J. and N{\'u}{\~n}ez-Y{\'e}pez, H. N. and Salas-Brito, A. L.},
  title   = {Solution of the Dirac Oscillator},
  journal = {Phys. Rev. Lett.},
  volume  = {64},
  pages   = {1643--1646},
  year    = {1990},
  doi     = {10.1103/PhysRevLett.64.1643}
}

@article{Berche2012ClassicalYang,
  author  = {Berche, Bertrand and Medina, Ernesto},
  title   = {Classical Yang--Mills Theory in Condensed Matter Physics},
  journal = {Eur. J. Phys.},
  volume  = {34},
  number  = {1},
  pages   = {161--180},
  year    = {2012},
  doi     = {10.1088/0143-0807/34/1/161}
}

@article{Bermudez2007MappingDirac,
  author  = {Bermudez, A. and Mart{\'i}n-Delgado, M. A. and Solano, E.},
  title   = {Exact Mapping of the $(2+1)$ Dirac Oscillator onto the Jaynes--Cummings Model: Ion-Trap Experimental Proposal},
  journal = {Phys. Rev. A},
  volume  = {76},
  number  = {4},
  pages   = {041801},
  year    = {2007},
  doi     = {10.1103/PhysRevA.76.041801}
}

@article{Bermudez2008ChiralityTransition,
  author  = {Berm{\'u}dez, A. and Mart{\'i}n-Delgado, M. A. and Luis, A.},
  title   = {Chirality quantum phase transition in the Dirac oscillator},
  journal = {Phys. Rev. A},
  volume  = {77},
  pages   = {063815},
  year    = {2008},
  doi     = {10.1103/PhysRevA.77.063815}
}

@article{Castroneto2009ElectronicProperties,
  author  = {Castro Neto, A. H. and Guinea, F. and Peres, N. M. R. and Novoselov, K. S. and Geim, A. K.},
  title   = {The Electronic Properties of Graphene},
  journal = {Rev. Mod. Phys.},
  volume  = {81},
  number  = {1},
  pages   = {109--162},
  year    = {2009},
  doi     = {10.1103/RevModPhys.81.109}
}

@article{Dalibard2011NeutralAtoms,
  author  = {Dalibard, Jean and Gerbier, Fabrice and Juzeli{\=u}nas, Gediminas and {\"O}hberg, Patrik},
  title   = {Colloquium: Artificial Gauge Potentials for Neutral Atoms},
  journal = {Rev. Mod. Phys.},
  volume  = {83},
  number  = {4},
  pages   = {1523--1543},
  year    = {2011},
  doi     = {10.1103/RevModPhys.83.1523}
}

@article{Dossa2020ParticleCarrying,
  author  = {Dossa, Finagnon A. and Avossevou, Gabriel Y. H.},
  title   = {Relativistic Dynamics for a Particle Carrying a Non-Abelian Charge in a Non-Abelian Background Electromagnetic Field},
  journal = {J. Math. Phys.},
  volume  = {61},
  number  = {2},
  pages   = {022302},
  year    = {2020},
  doi     = {10.1063/1.5123595}
}

@article{Dossa2020PauliHamiltonian,
  author  = {Dossa, Finagnon Anselme},
  title   = {Pauli Hamiltonian for a Spin One-Half Particle Carrying a Non-Abelian Charge in the Presence of Non-Abelian External Fields},
  journal = {Europhys. Lett.},
  volume  = {131},
  number  = {2},
  pages   = {21002},
  year    = {2020},
  doi     = {10.1209/0295-5075/131/21002}
}

@article{Francovillafane2013ExperimentalRealization,
  author  = {Franco-Villafa{\~n}e, J. A. and Sadurn{\'i}, E. and Barkhofen, S. and Kuhl, U. and Mortessagne, F. and Seligman, T. H.},
  title   = {First Experimental Realization of the Dirac Oscillator},
  journal = {Phys. Rev. Lett.},
  volume  = {111},
  number  = {17},
  pages   = {170405},
  year    = {2013},
  doi     = {10.1103/PhysRevLett.111.170405}
}

@article{Frohlich1993InvarianceCurrent,
  author  = {Fr{\"o}hlich, J{\"u}rg and Studer, U. M.},
  title   = {Gauge Invariance and Current Algebra in Nonrelativistic Many-Body Theory},
  journal = {Rev. Mod. Phys.},
  volume  = {65},
  number  = {3},
  pages   = {733--802},
  year    = {1993},
  doi     = {10.1103/RevModPhys.65.733}
}

@article{Gerritsma2010SimulationDirac,
  author  = {Gerritsma, R. and Kirchmair, G. and Z{\"a}hringer, F. and Solano, E. and Blatt, R. and Roos, C. F.},
  title   = {Quantum Simulation of the Dirac Equation},
  journal = {Nature},
  volume  = {463},
  number  = {7277},
  pages   = {68--71},
  year    = {2010},
  doi     = {10.1038/nature08688}
}

@article{Ghosh2025LandauLevels,
  author  = {Ghosh, Aritra},
  title   = {Landau levels of a Dirac electron in graphene from non-uniform magnetic fields},
  journal = {Phys. Lett. A},
  volume  = {561},
  pages   = {130956},
  year    = {2025},
  doi     = {10.1016/j.physleta.2025.130956}
}

@article{Goldman2014FieldsUltracold,
  author  = {Goldman, N. and Juzeli{\=u}nas, G. and {\"O}hberg, P. and Spielman, I. B.},
  title   = {Light-Induced Gauge Fields for Ultracold Atoms},
  journal = {Rep. Prog. Phys.},
  volume  = {77},
  number  = {12},
  pages   = {126401},
  year    = {2014},
  doi     = {10.1088/0034-4885/77/12/126401}
}

@article{Gonzalez2016ConfiningRepulsive,
  author  = {Gonz{\'a}lez, J.},
  title   = {Confining and repulsive potentials from effective non-Abelian gauge fields in graphene bilayers},
  journal = {Phys. Rev. B},
  volume  = {94},
  pages   = {165401},
  year    = {2016},
  doi     = {10.1103/PhysRevB.94.165401}
}

@article{Ito1967ExampleDynamical,
  author  = {Ito, D. and Mori, K. and Carriere, E.},
  title   = {An Example of Dynamical Systems with Linear Trajectories},
  journal = {Nuovo Cim. A},
  volume  = {51},
  pages   = {1119--1121},
  year    = {1967},
  doi     = {10.1007/BF02721683}
}

@article{Jagannathan1990DiracOscillator,
  author  = {Jagannathan, R.},
  title   = {The Dirac Oscillator and Its Supersymmetric Structure},
  journal = {Phys. Rev. A},
  volume  = {42},
  pages   = {6674--6677},
  year    = {1990},
  doi     = {10.1103/PhysRevA.42.6674}
}

@misc{Mandal2010DiracOscillator,
  author  = {Mandal, Bhabani Prasad and Verma, Shweta},
  title   = {Dirac oscillator in an external magnetic field},
  journal = {Phys. Lett. A},
  year    = {2010},
  doi     = {10.1016/j.physleta.2009.12.048},
  url     = {https://doi.org/10.1016/j.physleta.2009.12.048},
  note    = {Physics Letters A 374, 1021--1023 (2010); DOI: 10.1016/j.physleta.2009.12.048}
}

@article{Moshinsky1989DiracOscillator,
  author  = {Moshinsky, Marcos and Szczepaniak, Arturo},
  title   = {The Dirac Oscillator},
  journal = {J. Phys. A: Math. Gen.},
  volume  = {22},
  number  = {17},
  pages   = {L817--L819},
  year    = {1989},
  doi     = {10.1088/0305-4470/22/17/002}
}

@article{Novoselov2005TwoDimensional,
  author  = {Novoselov, K. S. and Geim, A. K. and Morozov, S. V. and Jiang, D. and Katsnelson, M. I. and Grigorieva, I. V. and Dubonos, S. V. and Firsov, A. A.},
  title   = {Two-Dimensional Gas of Massless Dirac Fermions in Graphene},
  journal = {Nature},
  volume  = {438},
  number  = {7065},
  pages   = {197--200},
  year    = {2005},
  doi     = {10.1038/nature04233}
}

@article{Ramires2018ElectricallyTunable,
  author  = {Ramires, Aline and Lado, Jos{\'e} L.},
  title   = {Electrically Tunable Gauge Fields in Tiny-Angle Twisted Bilayer Graphene},
  journal = {Phys. Rev. Lett.},
  volume  = {121},
  pages   = {146801},
  year    = {2018},
  doi     = {10.1103/PhysRevLett.121.146801}
}

@book{Ryder1996QuantumField,
  author    = {Ryder, Lewis H.},
  title     = {Quantum Field Theory},
  year      = {1996},
  publisher = {Cambridge University Press},
  address   = {Cambridge},
  edition   = {2},
  doi       = {10.1017/CBO9780511813900},
  isbn      = {9780521478144}
}

@article{Sanjose2012GrapheneBilayers,
  author  = {San-Jose, Pablo and Gonz{\'a}lez, Jos{\'e} and Guinea, Francisco},
  title   = {Non-Abelian Gauge Potentials in Graphene Bilayers},
  journal = {Phys. Rev. Lett.},
  volume  = {108},
  number  = {21},
  pages   = {216802},
  year    = {2012},
  doi     = {10.1103/PhysRevLett.108.216802}
}

@article{Tan2011SpinFields,
  author  = {Tan, S. G. and Jalil, M. B. A. and Fujita, T. and Liu, X.-J.},
  title   = {Spin Dynamics under Local Gauge Fields in Chiral Spin--Orbit Coupling Systems},
  journal = {Ann. Phys.},
  volume  = {326},
  number  = {2},
  pages   = {207--215},
  year    = {2011},
  doi     = {10.1016/j.aop.2010.11.014}
}

@article{Tan2020YangMills,
  author  = {Tan, Seng Ghee and Chen, Son-Hsien and Ho, Cong Son and Huang, Che-Chun and Jalil, Mansoor B. A. and Chang, Ching Ray and Murakami, Shuichi},
  title   = {Yang--Mills Physics in Spintronics},
  journal = {Phys. Rep.},
  volume  = {882},
  pages   = {1--36},
  year    = {2020},
  doi     = {10.1016/j.physrep.2020.08.002}
}

@article{Vozmediano2010FieldsGraphene,
  author  = {Vozmediano, M. A. H. and Katsnelson, M. I. and Guinea, F.},
  title   = {Gauge Fields in Graphene},
  journal = {Phys. Rep.},
  volume  = {496},
  number  = {4--5},
  pages   = {109--148},
  year    = {2010},
  doi     = {10.1016/j.physrep.2010.07.003}
}

@article{Wu1975NonintegrableFields,
  author  = {Wu, T. T. and Yang, C. N.},
  title   = {Concept of Nonintegrable Phase Factors and Global Formulation of Gauge Fields},
  journal = {Phys. Rev. D},
  volume  = {12},
  number  = {12},
  pages   = {3845--3857},
  year    = {1975},
  doi     = {10.1103/PhysRevD.12.3845}
}

@article{Yang1954ConservationIsotopic,
  author  = {Yang, C. N. and Mills, Robert L.},
  title   = {Conservation of Isotopic Spin and Isotopic Gauge Invariance},
  journal = {Phys. Rev.},
  volume  = {96},
  number  = {1},
  pages   = {191--195},
  year    = {1954},
  doi     = {10.1103/PhysRev.96.191}
}

@misc{Ariel2012EffectiveMassEnergyMass,
  author        = {Ariel, Viktor},
  title         = {Effective Mass and Energy-Mass Relationship},
  year          = {2012},
  doi           = {10.48550/arXiv.1205.3995},
  url           = {https://arxiv.org/abs/1205.3995},
  eprint        = {1205.3995},
  archiveprefix = {arXiv},
  primaryclass  = {physics.gen-ph}
}

@misc{ArielNatan2012ElectronEffectiveMassGraphene,
  author        = {Ariel, Viktor and Natan, Amir},
  title         = {Electron Effective Mass in Graphene},
  year          = {2012},
  doi           = {10.48550/arXiv.1206.6100},
  url           = {https://arxiv.org/abs/1206.6100},
  eprint        = {1206.6100},
  archiveprefix = {arXiv},
  primaryclass  = {physics.gen-ph}
}

@article{JaynesCummings1963BeamMaser,
  author  = {Jaynes, E. T. and Cummings, F. W.},
  title   = {Comparison of Quantum and Semiclassical Radiation Theories with Application to the Beam Maser},
  journal = {Proc. IEEE},
  volume  = {51},
  number  = {1},
  pages   = {89--109},
  year    = {1963},
  doi     = {10.1109/PROC.1963.1664}
}

@article{Knight2024JOSAB,
  author  = {Knight, P. L.},
  journal = {J. Opt. Soc. Am. B},
  volume  = {41},
  pages   = {C91},
  year    = {2024},
  doi     = {10.1364/JOSAB.524095},
}

@article{ShoreKnight1993JaynesCummings,
  author  = {Shore, B. W. and Knight, P. L.},
  title   = {The Jaynes-Cummings Model},
  journal = {J. Mod. Opt.},
  volume  = {40},
  number  = {7},
  pages   = {1195--1238},
  year    = {1993},
  doi     = {10.1080/09500349314551321}
}

@misc{Boumali2015ThermodynamicGraphene,
  author  = {Boumali, Abdelmalek},
  title   = {Thermodynamic Properties of the Graphene in a Magnetic Field via the Two-Dimensional Dirac Oscillator},
  journal = {Phys. Scr.},
  year    = {2015},
  doi     = {10.1088/0031-8949/90/4/045702},
  url     = {https://doi.org/10.1088/0031-8949/90/4/045702},
  note    = {Physica Scripta 90, 045702 (2015); DOI: 10.1088/0031-8949/90/4/045702}
}

@article{Zhang2025ExperimentalSimulationDirac,
  author  = {Zhang, Yu and Xu, Jianwen and Ma, Zhuang and Zheng, Wen and Lan, Dong and Tan, Xinsheng and Yu, Yang},
  title   = {Experimental Simulation of Dirac Equation in Superconducting Qubits},
  journal = {Commun. Phys.},
  volume  = {8},
  pages   = {248},
  year    = {2025},
  doi     = {10.1038/s42005-025-02112-2}
}

@article{Jiang2007InfraredLandau,
  author  = {Jiang, Z. and Henriksen, E. A. and Tung, L. C. and Wang, Y.-J. and Schwartz, M. E. and Han, M. Y. and Kim, P. and Stormer, H. L.},
  title   = {Infrared Spectroscopy of Landau Levels of Graphene},
  journal = {Phys. Rev. Lett.},
  volume  = {98},
  number  = {19},
  pages   = {197403},
  year    = {2007},
  doi     = {10.1103/PhysRevLett.98.197403}
}

@article{MandalRai2012NoncommutativeExternalMagnetic,
  author  = {Mandal, Bhabani Prasad and Rai, S. K.},
  title   = {Noncommutative Dirac Oscillator in an External Magnetic Field},
  journal = {Phys. Lett. A},
  volume  = {376},
  number  = {36},
  pages   = {2467--2470},
  year    = {2012},
  doi     = {10.1016/j.physleta.2012.07.001},
}

@misc{QuimbayStrange2013GrapheneDO,
  author        = {C. J. Quimbay and P. Strange},
  title         = {Graphene physics via the Dirac oscillator in (2+1) dimensions},
  year          = {2013},
  doi           = {10.48550/arXiv.1311.2021},
  url           = {https://doi.org/10.48550/arXiv.1311.2021},
  eprint        = {1311.2021},
  archiveprefix = {arXiv},
  primaryclass  = {cond-mat.mes-hall},
  note          = {arXiv:1311.2021; DOI: 10.48550/arXiv.1311.2021}
}

\end{document}